\pdfoutput=1
\documentclass[11pt]{article}

\usepackage[]{acl}

\usepackage{times}
\usepackage{latexsym}

\usepackage[T1]{fontenc}
\usepackage[utf8]{inputenc}

\usepackage{microtype}

\usepackage{hyperref}
\usepackage{enumitem}

\title{Fast and Light-Weight Answer Text Retrieval in Dialogue Systems}

\author{Hui Wan \\
  IBM Research AI \\
  \texttt{hwan@us.ibm.com} \And
Siva Sankalp Patel \\
  IBM Research AI \\
  \texttt{siva.sankalp.patel@ibm.com} \AND
J. William Murdock \\
  IBM Watson \\
  \texttt{murdockj@us.ibm.com} 
  \And
Saloni Potdar \\
  IBM Watson \\
  \texttt{potdars@us.ibm.com} \And
Sachindra Joshi \\
  IBM Research AI \\
  \texttt{jsachind@in.ibm.com} 
  }

\usepackage[colorinlistoftodos,prependcaption,textsize=tiny]{todonotes}

\begin{document}
\maketitle
\begin{abstract}

Dialogue systems can benefit from being able to search through a corpus of text to find information relevant to user requests, especially when encountering a request for which no manually curated response is available.  The state-of-the-art technology for neural dense retrieval or re-ranking involves deep learning models with hundreds of millions of parameters.  However, it is difficult and expensive to get such models to operate at an industrial scale, especially for cloud services that often need to support a big number of individually customized dialogue systems, each with its own text corpus.  We report our work on enabling advanced neural dense retrieval systems to operate effectively at scale on relatively inexpensive hardware.  We compare with leading alternative industrial solutions and show that we can provide a solution that is effective, fast, and cost-efficient. 

\end{abstract}

\section{Introduction}

Dialogue systems such as Amazon Lex, IBM Watson Assistant, or Microsoft Azure Bot Service operate mainly through intent detection.  A subject matter expert (SME) creates a dialogue system by defining a fixed set of intents that a user might have and provides scripted responses for each of them. Machine learning models are adopted to identify the user intent and route to the corresponding dialogue nodes and responses. It usually takes a considerable amount of human curated data to train an intent detection model. Adding features or content to a dialogue system would require adding new intents and training the model all over again.

To alleviate such limitations, an alternative approach to enabling the same user experience is to have a system automatically search through a corpus of text to find relevant responses to each user request.  One motivation behind this approach is to replace the intent detection, so to make it flexible, quicker, and easier to set up and maintain a dialogue system, because the SME does not need to enumerate all the intents they expect a user to have. Applying text retrieval in such a system can also complement intent detection: intent detection can handle the anticipated user needs and text search can handle unanticipated requests.  In either case, the value of the text retrieval depends critically on how accurate it is. Another big advantage of the text retrieval approach is that it could provide reasonable accuracy even when there is little or no labeled training data.

A popular line of text retrieval methods is matching sparse terms and weighing those matches by how frequent they are in the document being found and how infrequent they are in the corpus.  For example, BM25~\cite{bm25} is an extremely popular algorithm of this sort that provides an excellent balance between accuracy and computational cost. However, in the recent years, research has shown that neural network solutions can provide superior accuracy to sparse term matching approaches like BM25.  In particular, neural dense retrieval approaches such as DPR~\cite{karpukhin-etal-2020-dense} and  
ColBERT~\cite{khattab2020colbert, khattab2020relevance} have achieved outstanding results in retrieval and re-ranking even at zero-shot setting, and further boosted accuracy when in-domain training data is available.

Neural dense retrievers achieve high accuracy but usually involve models with hundreds of
millions of parameters and require long training time. However, in real-world scenarios,
a cloud service sometimes supports many different deployed dialogue applications at the same time, hence needs to be able to process requests for all of those applications at the same time.  This can be extremely expensive if each application has a model that demands an enormous amount of memory and/or processing power when handling requests.  A practical system needs to be able to balance the benefits of a sophisticated model with the costs of running it. Furthermore, dialogue system administrators want to be able to add training data to an existing, deployed system and start getting improved results quickly.

We explore various approaches to addressing these requirements, including scaling techniques such as distilled encoders and dimension reduction,
self-directed iterative learning and asynchronous learning. We conduct thorough experiments on our datasets to benchmark these approaches, and show that we have emerging technology that achieves accuracy that is competitive with state-of-the-art research solutions with substantially less expensive resource requirements. We made public the ColBERT code with our improvement \footnote{\url{http://github.com/IBM/ColBERT-practical}} to enable application to our use cases.

\section{Related Work}

In Information Retrieval (IR), popular relevancy algorithms such as TF-IDF and BM25 \cite{bm25} match keywords with an inverted index and compute relevancy using heuristic functions.
Together with pre-processing methods such as stemming and removal of curated stop words, sparse-term-based retrieval works fairly well without training, and is widely adopted in real world applications.

Dense passage retrieval~\cite{karpukhin-etal-2020-dense, khattab2020colbert, khattab2020relevance, xiong2021ance, luan-etal-2021-sparse, santhanam2021colbertv2} has gained a lot of attention lately with applications extending beyond retrieval tasks into areas including open-domain question answering, language model pre-training, fact checking, dialogue generation
(e.g., RAG~\cite{lewis2020retrieval}, 
REALM~\cite{guu2020realm},
MultiDPR~\cite{Maillard2021multidpr},
KILT~\cite{petroni2021kilt},
ConvDR~\cite{yu2021fewshot}, RocketQA~\cite{qu-etal-2021-rocketqa}).
In dense passage retrieval, the query $q$ and each passage $p$ are separately encoded
into dense vectors, and relevance is modeled via similarity functions such as dot-product.
%In single vector retrieval models such as DPR~\cite{karpukhin-etal-2020-dense} and BERT Siamese/Dual Encoder ~\cite{luan-etal-2021-sparse}, the query and passages are separately encoded into single vectors, models are trained with the objective of mapping the relevant passage close to the query, and pushing the irrelevant passage far away from the query. 
%During inference time, ANN (approximate nearest neighbor) search libraries such as FAISS~\cite{2017faiss} are then used to search directly  for the passage vectors closest to the query vector.
%In late interaction models such as ColBERT~\cite{khattab2020colbert, khattab2020relevance, santhanam2021colbertv2},the query and passages are separately encoded to obtain query token vectors and passage token vectors. The sum of maximum-similarity (SumMaxSim) scores to query vectors are used to model the relevance of passages. During training, models are trained with the objective of maximizing the SumMaxSim scores of relevant passages and minimizing those of irrelevant passages.
%During inference time, the passage tokens closest to query tokens are fetched, and the relevant passages are ranked based on the SumMaxSim scores.
Recent works improve efficiency and effectiveness of single-vector dense retrieval systems, including 
model distillation~\cite{hofstaetter2020_crossarchitecture_kd, lin-etal-2021-batch}, hard negative sampling~\cite{xiong2021ance, zhan-2021-hard-neg}, etc..
% Learning to retrieve: How to train a dense retrieval model effectively and efficiently

Another line of related work is cross-encoder document ranking~\cite{MacAvaney2019sigir, Dai2019sigir, nogueira2019passage}. 
Query–document pairs are concatenated and sent through Transformer-based encoders,
 an additional layer on top of the encoded representation is adopted to produce a
relevance score of the document to the query, which is then used for ranking.
%Because of the cross-encoder, this approach delivers good accuracy while being much more computationally expensive than the other approaches.

\citet{arora-etal-2020-hint3} and \citet{qi-etal-2021-benchmarking} benchmark intent detection models on intent detection datasets such as CLINC150~\cite{larson-etal-2019-evaluation}  where sufficient training examples exist for each intent. On the other hand, our use case focuses on the scenarios where answer text is available but training examples are insufficient.

\section{Task and Baselines}
\label{sec-task}

The task we are dealing with is a real-world use case of answer text retrieval in an FAQ dialogue system. 

Formally, we have a corpus $P$ of answer text snippets (passages). For each answer text passage $p$ in $P$, we have a limited number of associated example queries $Q_p$. The system is expected to retrieve the most relevant answer text passage for each incoming user query $q$. It needs to deliver a good latency, and work well when the size of $Q_p$ is small, i.e., when there are not many training examples available. Most importantly, the resource consumption must be kept low. 
 
To address the use case, we start with two leading industrial solutions as baselines:
\begin{itemize}[style=unboxed,leftmargin=0cm]
    \item%[Neural feature classifier] 
    One approach is to map each answer text $p$ as a class $c_p$, and train a classifier 
on $\{(c_p, q_p)$ for each $p$ and each $q_p$ in $Q_p\}$
to predict the incoming queries. With the recently ubiquitous large pre-trained language models such as BERT~\cite{devlin2019bert} and RoBERTa~\cite{roberta2019}, classifiers equipped with both hand-crafted features and neural embedding features are very powerful and deliver decent predictions when there are enough training examples. However, obtaining large amounts of high-quality training data is expensive.  Often there is little or no training data. 

    \item%[BM25] 
    Sparse-term-based retrieval (e.g., BM25) on the answer text  is another natural approach to address the task without 
the demand for training data. It has the advantage of having minimal resources requirement. On the other hand, it could not well leverage training data when it is available.

\end{itemize}

The two aforementioned approaches each have their own strength. The classifier approach leverages query examples and machine learning, while the sparse-term-based retrieval approach utilizes answer text but not query examples, and does not involve training. We seek to get the benefits from both approaches.  One option is to capture the cross-attention between query $q$ and each candidate passage $p$ by feeding $\langle q, p\rangle$ pair to a Transformer-based encoder and learn over the encoded output~\cite{MacAvaney2019sigir, Dai2019sigir, nogueira2019passage}. However, due to the need to cross-encode the incoming query together with each passage, this approach requires more computation by orders of magnitude and is not practical for our task setting.

Dense passage retrieval methods ~\cite{karpukhin-etal-2020-dense,khattab2020colbert, santhanam2021colbertv2, luan-etal-2021-sparse, Humeau2020Poly-encoders, Macavaney2020PreTTR, xiong2021ance} 
have gained a lot of attention lately and achieved state of the art results on various retrieval and ranking datasets.
Dense retrievers are efficient compared to other neural methods such as transformer-based cross-encoder models: passages are encoded and indexed offline, 
at inference time only the query needs to be encoded once; also they leverage ANN (approximate nearest neighbor) algorithms to efficiently search for relevant  dense vectors.
Dense retrievers are effective compared to traditional sparse-term-based IR methods such as BM25: They are not restricted by rigid keyword matching; They use transformers to encode both the queries and the passages, and benefit from transfer learning from large retrieval/re-ranking datasets.
Being effective and efficient, neural dense retrievers make an ideal solution for our task setting and requirements.

% The original baselines were moved to experiment settings.

\section{Approach}
\label{sec-approach}
 We first briefly overview the work in neural dense retrieval and talk about the gaps from practical usage in Section~\ref{sec-nir}.
In the remainder of Section~\ref{sec-approach}, we explain our efforts applying dense passage retrieval to the task and further reducing response time, memory footprint, and training time. 

\subsection{Neural Dense Retrieval Preliminaries}
\label{sec-nir}

%Dense passage retrieval methods such as DPR~\cite{karpukhin-etal-2020-dense} and ColBERT~\cite{khattab2020colbert} have gained a lot of attention lately and achieved state of the art results on various retrieval and ranking datasets. 
In dense passage retrieval, $q$ and $p$ are separately encoded.  All the passages can be encoded and indexed offline. During inference time, only the query needs to be encoded; ANN (approximate nearest neighbor) search libraries such as FAISS~\cite{2017faiss} are used to efficiently search for the most relevant passage. 
%These traits make dense passage retrieval a solution matching our task setting and requirements.

% Introduce briefly DPR and ColBERT...
In single-vector retrieval models such as DPR~\cite{karpukhin-etal-2020-dense} and BERT Siamese/Dual Encoder ~\cite{luan-etal-2021-sparse}, the query and passages are separately encoded into single vectors, 
models are trained with the objective of mapping the relevant passage vector close to the query vector, and pushing the irrelevant passage vectors far away from the query vector. 
During inference time, ANN search is used to retrieve directly for the passage vectors closest to the query vector. Several other systems leverage multi-vector
representations and attention-based re-ranking, including Poly-encoders~\cite{Humeau2020Poly-encoders}, PreTTR~\cite{Macavaney2020PreTTR}, etc..

In late interaction models such as
ColBERT~\cite{khattab2020colbert, khattab2020relevance, santhanam2021colbertv2},
the query and passages are separately encoded to obtain query token vectors and passage token vectors.
These models adopt token-decomposed scoring, e.g. the sum of maximum-similarity (SumMaxSim) scores to query vectors are used to model the relevance of passages.
During training, models are trained with the objective of maximizing the SumMaxSim scores of relevant passage and minimizing those of irrelevant passages.
During inference time, the passage tokens closest to query tokens are fetched, and then the relevant passages are re-ranked based on the SumMaxSim scores.

We experimented with two of the most popular dense retrieval models, DPR and ColBERT. As effective as they are, they still consume more computing resources and take longer response time than required in our real-world use case of hosting thousands of customized systems. Also, in our use case, dialogue system administrators want to reduce the time to fine-tune neural retrieval models on custom training data.

%We will compare the results in Section~\ref{sec-results}.

\subsection{Dense Retrieval Scaled for Practical Usage}
\label{sec-scale}

%We refer readers to the original papers for details.
For practical usage we implemented improvement features into ColBERT code
\footnote{We made public our improvement on ColBERT at \url{http://github.com/IBM/ColBERT-practical}.}
: 1) for encoder, add flexible accommodation for various transformer types and models in the Huggingface model hub; 2) new improved batcher and training loop logic by epochs, flexible shuffling and checkpoint saving.

We benchmark DPR and ColBERT on our datasets, and experiment reducing response time and memory footprint at retrieval time as follows.

\begin{description}[style=unboxed,leftmargin=0cm]
%\noindent\textbf{Distilled transformer encoder}
\item[Distilled transformer encoder]
We pre-train ColBERT model on the Natural Questions (NQ) dataset~\cite{naturalquestions} from multiple small-size or distilled transformers models including Electra~\cite{clark2020electra}, TinyBERT~\cite{tinybert2020jiao}, DistilBERT~\cite{sanh2019distilbert} and DistilRoBERTa. After comparing the memory footprint, the retrieval time, and the retrieval accuracy, we chose to use TinyBERT~\cite{tinybert2020jiao} with 4 layers and 312 hidden dimensions\footnote{\url{https://huggingface.co/huawei-noah/TinyBERT_General_4L_312D}}.
    
%\noindent\textbf{Dimension reduction} 
\item[Dimension reduction]
\cite{khattab2020colbert, santhanam2021colbertv2} showed that a ColBERT model with quantized and reduced-dimension vectors could perform comparably to the standard model on big retrieval/ranking benchmarks  while greatly reducing the space requirement for saving the final representations. For our use case on the small retrieval datasets, we explored using smaller dimensions for the vector representations in ColBERT. In our experiments, however, reduced dimension models yield much lower accuracy.
    
% \noindent\textbf{Shorter query length} 
\item[Shorter query length]
We decrease the maximum query length in DPR from 256 to 32, reducing the response time of DPR by ~80\%. As this length still fits the majority of the queries in our task setting, the effect to accuracy is very tiny and could be neglected.

\end{description}

\subsection{Self-directed Iterative Learning }
\label{sec-iter}
Dense retrieval training data consists of $\langle q, p^+, p^- \rangle$ triples, where $q$ is the query, $p^+$ is a positive (relevant) passage, and $p^-$ is a negative (irrelevant) passage. Dense neural retrieval models learn from such triples to effectively map query token representations and relevant answer text token representations together, and push irrelevant (token) representations away.
While forming training triples, one straightforward way is using all the negative passages to make sure not missing any useful training data.
However, this results in long training time.
%This leads to some too easy negative passages being included in training, and unnecessarily increased training time. 
Sampling from BM25 top ranked passages is a widely used approach to select negative passages. However, this introduces a data bias and limit the model's learning ability~\cite{luan-etal-2021-sparse}.
An alternative approach is to choose negatives passages from those highly ranked by the model from the previous training iteration.  This allows each iteration of the training to learn from negative examples for which the previous model did not do well \cite{2015hardnegative, 2017samplingMatters}. 

\begin{figure}
\centering
\includegraphics[width=3.1in]{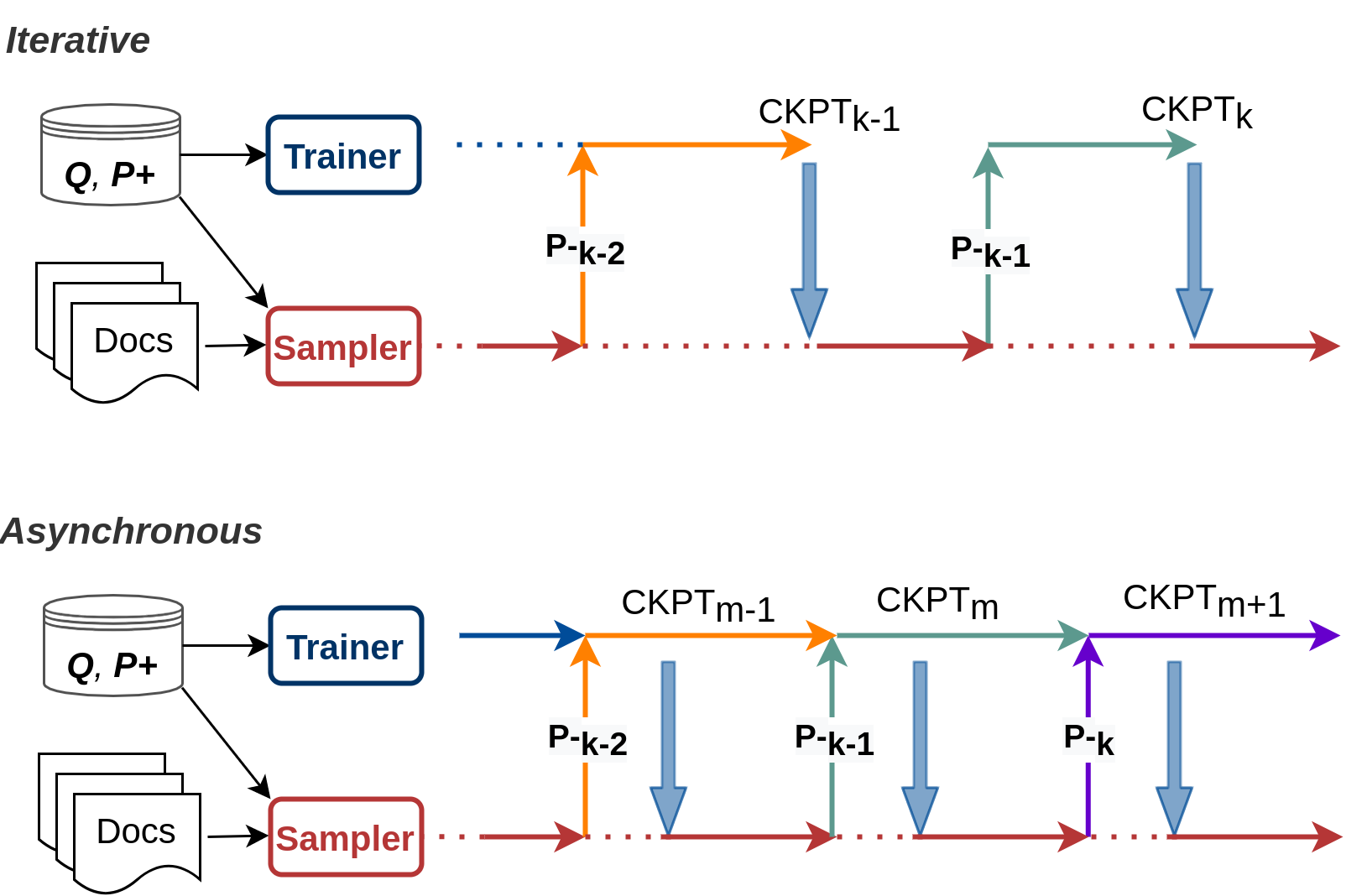}
\caption{Iterative learning strategy~\ref{sec-iter} and asynchronous learning strategy~\ref{sec-asyn}.}
\label{fig:iter-asyn}
\vspace{-0.15in}
\end{figure}

To be more specific, given a trained ColBERT model $\tt{CKPT}$, we take a query $q$ from training data, and get $\tt{CKPT}$'s top $m$ ranked passages ($p_1, ..., p_{m}$) for $q$, suppose the positive passage is $p_{i}$, we take each negative passage ranked higher than $p_i$ to form the new batch of training triples
$\langle q, p_i, p_1\rangle,... \langle q, p_i, p_{i-1}\rangle $. When $i=1$, i.e., the model gave the right prediction, we still include several randomly sampled triples, so as to avoid over-fitting on a few difficult queries.

With the self-directed triple curation, we explore an iterative learning strategy as illustrated in Figure~\ref{fig:iter-asyn}.
In each iteration, the Sampler module and the Trainer module work together as follows. 
In each iteration, first, Sampler uses a recently trained model checkpoint $\tt{CKPT}_{k-1}$ to update the representation of documents in the corpus and refresh the ANN index, then from the refreshed ANN index  fetch the top ranked negatives $P^-_{k-1}$ for training queries $Q$ to produce training triples together with $P^+$.
Then, Trainer uses the triples generated by Sampler to train a new model checkpoint $\tt{CKPT}_{k}$.
%In the next iteration, Sampler takes this newly trained model checkpoint to produce a new batch of training triples, which in turn will be consumed by Trainer. 
In each iteration, only the negative examples that are ``hard'' for the current model are used to form the training triples, thus we achieve effective and focused training with reduced time.
 Note that similar strategy was adopted by \citet{khattab2020relevance} by training two more stages after the initial ColBERT model. We make the further exploration by automatically continuing the iterations until the model reached certain accuracy on training queries.

\subsection{Asynchronous Learning}
\label{sec-asyn}
During the iterative learning in Section~\ref{sec-iter}, the Trainer and Sampler wait for each other's output to proceed to next round.  This causes overhead and wasted resources. To alleviate that, we adopt the asynchronous learning approach as described in ANCE~\cite{xiong2021ance} and 
let the Trainer and Sampler work asynchronously without waiting on each other, as depicted in Figure~\ref{fig:iter-asyn}. 
To be specific, while Sampler is curating the new batch, the Trainer does not wait but continues training on the old batch of training triples. After generating a batch of training triples, the Sampler always fetches the latest model checkpoint and starts creating a new batch.
Note that the implementation in ANCE~\cite{xiong2021ance} is on BERT Siamese/Dual Encoder~\cite{luan-etal-2021-sparse}.  As far as we know, our implementation is the first on ColBERT model.

\subsection{Ensemble}
\label{sec-ensemble}

With the scaling efforts in Section~\ref{sec-scale}, we achieve a neural dense retriever with a latency comparable to neural-embedding-based SVM  and BM25.
This makes it practical to ensemble the two systems with the neural dense retrieval system.
We ensemble a neural-embedding-based SVM classifier and neural retrieval in scenarios where training data is available,
and ensemble BM25 and neural retrieval in scenarios where training data is unavailable.

\section{Experiments}
\label{sec-exp}

\subsection{Datasets}
\label{sec-dataset}

For our experiments, we obtain datasets from real-world dialogue systems.
We create datasets from an HR policy FAQ bot (denoted by HRFAQ) and a medical group portal FAQ bot (denoted by MEDFAQ), both in English.  
Each dialogue system dataset consists of intents, intent examples, dialogue node graphs and response texts created by subject-matter experts. For each dataset, we created a test set of queries and ground truth responses by sampling the real-world  chat logs from the deployed dialogue system.
The task is measured by Match@1 score in results tables, which is the percentage of test queries for which the top system result is correct.
Table~\ref{table-dataset} shows the dataset statistics.
Note that the datasets are not big and the queries are generally short.  The challenge in scaling comes mainly from trying to support many such systems at once in the same cloud.

\begin{table}
	\centering
	\small
	\begin{tabular}{lcc}
		\hline
		 \textbf{Dataset} & \textbf{HRFAQ}  & \textbf{MEDFAQ}   \\ 
		 \hline\hline
		\# docs   & 186 & 87 \\
		\# words / doc  & 35.4 & 31.4 \\

	    \# training queries & 5433 & 862 \\
	    \# words / train query  & 8.8 & 4.9 \\

        \# test queries   & 1174 & 462 \\
        \# words / test queries   & 6.6 & 4.5 \\

		\hline
	\end{tabular}
	\vspace{3pt}
\caption{\label{table-dataset}
		Dataset statistics.
	}
\end{table}

\subsection{Experimental Settings}
\label{sec-exp-settings}

For the sparse-term-based retrieval baseline, we use BM25 \cite{bm25} as implemented in ElasticSearch\footnote{\url{http://www.elastic.co/elasticsearch/}}, with lower-casing, stemming and stop-word removal.

For the neural-embedding-based classifier, we train a one vs all SVM classifier
with sophisticated pre-processing, hand-crafted n-gram features, and neural word/sentence embeddings based on Transformers with 512-dimension vectors\footnote{We refrain from giving more details because this is a commercial product.}. We also train a classifier with answer text added as training queries, denoted by `` NSVM w/ text'', as opposed to ``NSVM'' which does not use answer text hence has no 0-shot numbers.

For DPR experiments, we use the Facebook research DPR repository\footnote{\url{http://github.com/facebookresearch/DPR}}. %Hyper-parameters ***.
The DPR full model before fine-tuning is downloaded from the DPR repository (March 2021 release). The DPR$_{tiny}$ model before fine-tuning is pre-trained on the triples created from Natural Questions (NQ) dataset~\cite{naturalquestions}, also obtained from the same repository.  ``DPR(S)'' stands for shorter query setting.

For ColBERT experiments, our code is built on top of the v0.2 version of ColBERT code\footnote{\url{http://github.com/stanford-futuredata/ColBERT}}, which is in PyTorch and uses Huggingface Transformers\footnote{\url{http://github.com/huggingface/transformers}}.
We implemented the code for iterative learning and asynchronous learning in PyTorch.
For real-world usage we also implemented improvement features into ColBERT code as described in Section~\ref{sec-scale}.
%%, including ***.

The ColBERT full model before fine-tuning is provided by the authors of ColBERT. 
The ColBERT$_{tiny}$ model before fine-tuning  is pre-trained on triples created from Natural Questions (NQ) dataset~\cite{naturalquestions} as specified in ColBERT~\cite{khattab2020relevance}.

%CPU environment in experiments.
For CPU environment inferencing, all models and data/indices reside locally on a CPU machine with four Intel® Core™ i7-8650U CPUs. 
Neural models are trained on a single NVIDIA V100 GPU in a computing cluster environment unless otherwise stated.

Hyper-parameters and other detailed settings are included in Appendix.

\subsection{Experiments and Results}
\label{sec-results}

%%%%%%%%%%%%%%%%%%%%%%%
\begin{description}[style=unboxed,leftmargin=0cm]
% \noindent\textbf{Resources Consumption}
\item[Resources Consumption]
Table~\ref{table-scale} compares the resource usage and response times of different systems during inference (retrieval). Full Neural models have high memory consumption and consume a lot of disk space because of millions of parameters in the neural networks. The smaller dense retrieval models, as scaled in Section~\ref{sec-scale}, are able to reduce both footprints and inference latency drastically.
%This shows that the scaling methods in Section~\ref{sec-scale} help greatly reduce the resources usage and inference time. 

\begin{table}
	\centering
	\small
	\begin{tabular}{lrrr}
		\hline
		\hline
		 {\textbf{System}} & \textbf{Size} & \textbf{Mem} & \textbf{Time}  \\ \hline
	{BM25}    & $-$ & $-$ & $4.6$ms  \\

	{NSVM}    & $1.1$G & $2.9$G & $10$ms  \\

	{DPR}    & $836$M & $2.5$G & $267$ms   \\

	{ColBERT}   & $419$M & $2.4$G & $59$ms  \\

	{DPR(S)}   & $836$M & $2.5$G  & $45$ms   \\

	{DPR$_{tiny}$(S)}   & $110$M & $0.6$G & $5$ms \\

	{ColBERT$_{tiny}$}   & $55$M & $1.7$G & $10$ms \\
		\hline
		\hline
	\end{tabular}
	\vspace{3pt}
\caption{\label{table-scale}
		Inference latency and resources usage of different systems on HRFAQ dataset in CPU environment. Latency is for single query and includes pre-processing time.
		%All models and data/indices reside locally on a CPU machine with four Intel® Core™ i7-8650U CPU. 
		%{*: Customizable ElasticSearch/Lucene configuration.}
		DPR and ColBERT model sizes do not include optimizer variables.
	}
\end{table}

%\noindent\textbf{Choosing Distilled Base Models}
\item[Choosing Distilled Base Models]
We conduct further benchmarking on ColBERT models based on different distilled language models\footnote{All models downloaded from Huggingface model hub \url{https://huggingface.co/models}.}
including
DistilBERT$_{base}$, DistilRoBERTa$_{base}$, Electra$_{small,discriminator}$, TinyBERT$_{4L-312}$ and TinyBERT$_{6L-768}$. We pre-train a ColBERT model from each of these transformer models, and test on the 0-shot setting of the HRFAQ dataset. An alternative approach would be to distill from fully trained ColBERT models using the corresponding distillation algorithms, which we leave for future work.
All models are pre-trained on the NQ dataset at a batch size of 192 for 40k steps, except Electra and DistilRoBERTa are trained for 80k steps because of their lower accuracy at 40k steps.
The results suggest that the general pre-training before ColBERT training does impact generalization performance of the ColBERT models.
Specifically, larger models, e.g., DistilRoBERTa$_{base}$, do not always result in better generalization, and starting from TinyBERT$_{4L-312}$ appears to be a good choice considering efficiency and accuracy.
%empirically outperforms six-layer TinyBERT. 
We use TinyBERT$_{4L-312}$ as the distilled base model in the remainder of the paper and denote it by ${tiny}$.
The full models trained from BERT$_{base}$ are sub-scripted by ${full}$.

\begin{table}
	\centering
	\small
	\begin{tabular}{lrrrr}
		\hline 
		\hline
		 {\textbf{System}} & \textbf{Size} & \textbf{Mem} & \textbf{Time} & M@1 \\ \hline
	DistilBERT            & $254$M & $2.3$G & $26$ms & 35.0 \\
	DistilRoBERTa           & $314$M & $3.8$G & $32$ms & 32.3 \\
	TinyBERT$_{6L-768}$   & $256$M & $2.1$G & $27$ms & 35.3 \\
	TinyBERT$_{4L-312}$   & $55$M & $1.7$G & $10$ms & 36.3 \\
	Electra               & $52$M & $1.7$G & $18$ms & 29.5 \\
		\hline
		\hline
	\end{tabular}
	\vspace{3pt}
\caption{\label{table-scale-2}
		Inference latency, resources usage, and accuracy of different ColBERT models on HRFAQ dataset in a CPU environment.
	}
\end{table}

%\noindent\textbf{Fine-tuning Accuracy} 
\item[Fine-tuning Accuracy]
Tables ~\ref{table-results-1} and \ref{table-results-2} show results or HRFAQ and MEDFAQ.
%For both datasets, the full ColBERT system is the best performing single system. 
ColBERT$_{full}$ is the most accurate single system especially in 0-shot setting, which is consistent with results from research papers.
With more training examples, DPR catches up in accuracy, showing that retrieval methods based on single vector similarity instead of token vector late interactions is at disadvantage transferring to 0-shot use cases, but performs nicely with some training examples.
It is worth noting that, ColBERT$_{tiny}$ shows only a small degradation from ColBERT$_{full}$ on HRFAQ, presenting a nice trade-off
between accuracy and efficiency in real-world industry use cases.  In MEDFAQ, there is a bigger drop in accuracy from ColBERT$_{full}$ to ColBERT$_{tiny}$.  This may be a result of MEDFAQ's vocabulary and content being more distant from the NQ data used for pre-training, since medical vocabulary tends to be highly specialized.  In 1-shot and 3-shot settings where the models are trained with 1 or 3 examples per answer, ColBERT$_{tiny}$ is more competitive for MEDFAQ.

\begin{table}
\centering
\small
\begin{tabular}{rlccc}
	\hline
	\hline
& \textbf{HRFAQ} & \multicolumn{1}{c}{\textbf{0-shot}} & \multicolumn{1}{c}{\textbf{1 ex/doc}} &  \multicolumn{1}{c}{\textbf{3 ex/doc}}   \\ \hline
%		\# & System & M@1    & M@1   & M@1    \\ \hline
	\hline
    1 &BM25   & 29.2 &  $-$ & $-$   \\
	2 &	NSVM  & $-$  & 23.2(4.4)  & 43.3(3.6)   \\
	3 &	NSVM w/ text  & 10.4 & 27.5(3.7) & 46.0(3.5)   \\
	4 &	{DPR$_{full}$} & 29.9  & 42.3(2.6)  & 53.5(2.2)    \\
	5 &{ColBERT$_{full}$} & 38.9 & 47.8(1.8)  & 53.6(2.3)  \\
	    \hline
	6 &	{DPR$_{tiny}$(S)} & 25.7  & 37.8(2.9) & 46.2(4.1)     \\
	7 &{ColBERT$_{tiny}$} & 36.3  & 42.4(1.7) & 50.7(2.0) \\
	8 & Ensemble(1,7) & \textbf{39.0}  & \textbf{47.4(1.8)}  & 53.4(2.2)  \\
	9 & Ensemble(3,7) & 30.4 & 45.0(2.3)  & \textbf{55.4(2.0)}  \\
	\hline
	\hline
\end{tabular}
	%\vspace{15pt}
\caption{\label{table-results-1}
Match@1 scores on HRFAQ test set.
For $k$ ex/doc experiments: we take 10 random seeds; for each random seed, sample $k$ training
queries per answer text, train a model; finally report avg(std) of the 10 models.
Scores in bold are best in efficient setting.
	}
\end{table}

\begin{table}
\centering
\small
\begin{tabular}{rlccc}
	\hline
	\hline
& \textbf{MEDFAQ} & \multicolumn{1}{c}{\textbf{0-shot}} & \multicolumn{1}{c}{\textbf{1 ex/doc}} &  \multicolumn{1}{c}{\textbf{3 ex/doc}}   \\ \hline
%		\# & System & M@1    & M@1   & M@1    \\ \hline
	\hline
    1 &BM25   & 25.1 &  $-$ & $-$   \\
	2 &	NSVM  & $-$  & 39.7(5.1)  &  60.0(6.4)  \\
	3 &	NSVM w/ text  & 22.5 & 41.4(4.9) & 58.6(4.7)    \\
	4 &	{DPR$_{full}$} & 37.0 & 58.5(3.7)  & 67.2(2.2)   \\
	5 &{ColBERT$_{full}$} & 45.2 & 57.6(2.5)  & 67.7(1.7)  \\
	    \hline
	6 &	{DPR$_{tiny}$(S)} & 25.5 & 44.7(5.0) & 56.9(4.1)     \\
	7 &{ColBERT$_{tiny}$} & 26.6   & 47.1(4.0)  & 60.4(4.4) \\
	8 & Ensemble(1,7)  & 28.6 & 47.5(4.2)  &  60.4(4.1)  \\
	9 & Ensemble(3,7) & \textbf{29.9}  & \textbf{51.3(7.0)}  & \textbf{63.3(5.0)}   \\
	\hline
	\hline
\end{tabular}
	%\vspace{15pt}
\caption{\label{table-results-2}
Match@1 scores on MEDFAQ test set.
Details same as Table~\ref{table-results-1}.
%For $k$ ex/doc experiments: we take 10 random seeds; for each random seed, sample $k$ trainingqueries per answer text, train a model; finally report avg(std) of the 10 models. Scores in bold are best in efficient setting.
	}
\end{table}

%%%%%%%%%%%%%%%%%%%%%%%
%\noindent\textbf{Ensembling}
\item[Ensembling]
We take a linear combination of 0-shot BM25 predictions and ColBERT$_{tiny}$ predictions with heuristic weight $0.3$:$1$,  
and a $10$:$1$ combination of SVM predictions and ColBERT$_{tiny}$ predictions, since the scores from the SVM classifier are in a higher magnitude. 
As shown in the second parts of Tables~\ref{table-results-1} and Table~\ref{table-results-2}, there is a nice boost from both systems being ensembled, showing ensembling to be a feasible and effective approach to further increase the accuracy.

%(To be added) A table comparing reduced dimension.
%\todo{Add a comparison table}

%%%%%%%
%\noindent\textbf{Self-guided Iterative / Asynchronous Learning}
\item[Self-guided Iterative / Asynchronous Learning]
Table~\ref{table-asyn} compares the retrieval results and training time efficiency of one-pass training with all negatives, 
one-pass training with BM25 guided negatives, 
iterative learning, and asynchronous learning. We use ColBERT$_{tiny}$ for this comparison.
For BM25 guided and iterative/asynchronous learning, negative examples are curated as described in Section~\ref{sec-iter}, from top 20 model-guided predictions.
Models are trained for 10 epochs in the one-pass experiments, and 5 rounds of 6 epochs each in the iterative and asynchronous learning experiments. The results demonstrate that, with iterative self-guided sampling of negative passages, ColBERT models can achieve results competitive to the models trained on complete data within ~20\% training time.
The M@1 score of \texttt{All neg} is slightly lower at 1-shot, likely due to the mismatch of randomly sampled training examples and the testset.

%We also include the chart of testset accuracy changing by time in different approaches. 
%(If have time, comparing different hyperparameters of negative sampling.)

\begin{table}
	\centering
	\small
	\begin{tabular}{lrcrc}
		\hline
		\hline
		 \textbf{HRFAQ} &\multicolumn{2}{c}{\textbf{1 ex/doc}} &  \multicolumn{2}{c}{\textbf{3 ex/doc}}   \\ \hline
		 ColBERT$_{tiny}$ & Time & M@1  & Time & M@1   \\ \hline
		\hline
	All neg  & 475s & 42.4(1.7)   & 1374s  & 50.7(2.0)   \\
   BM25 Guided &  37s & 37.1(1.7)  &  85s & 39.4(2.1)      \\
	Iterative  & 104s & 44.0(2.1)  & 226s  & 49.4(1.6)     \\
	Asynchronous  & 78s & 43.0(2.1)  & 200s & 49.3(1.9)    \\

		\hline
		\hline
	\end{tabular}
	
	\vspace{3pt}

\caption{\label{table-asyn}
		Training time and Match@1 scores of different training strategies.
		Scores avg(std) on 10 randomly sampled training sets.
		%The numbers are avg(std) of scores on 10 models trained on randomly sampled training sets. 
	}
\end{table}

%\noindent\textbf{Summary}  
\item[Summary]
Compared to BM25, dense passage retrieval has the advantage of utilizing small amount of training data and gaining a much higher accuracy; compared to neural classifiers, it works well with little or no training data and is flexible with changes in the prediction space.
Although with resources consumptions higher than BM25, dense passage retrieval with scaling techniques could deliver higher accuracy than BM25 and neural embedding based classifiers with similar latency, thus makes a great solution for our use case.

\end{description}

\section{Conclusion}
We report on our work on enabling advanced neural dense retrieval systems to operate effectively at scale on relatively inexpensive hardware. On our real-world use case and datasets from dialogue systems, we show that we can provide a solution that  achieves accuracy that is competitive with state-of-the-art research solutions with substantially less expensive resource requirements and shorter response time.
% Acknowledgement...

\bibliography{anthology,custom}

\begin{thebibliography}{33}
\expandafter\ifx\csname natexlab\endcsname\relax\def\natexlab#1{#1}\fi

\bibitem[{Arora et~al.(2020)Arora, Jain, Chaturvedi, and
  Modi}]{arora-etal-2020-hint3}
Gaurav Arora, Chirag Jain, Manas Chaturvedi, and Krupal Modi. 2020.
\newblock \href {https://doi.org/10.18653/v1/2020.insights-1.16} {{HINT}3:
  Raising the bar for intent detection in the wild}.
\newblock In \emph{Proceedings of the First Workshop on Insights from Negative
  Results in NLP}, pages 100--105, Online. Association for Computational
  Linguistics.

\bibitem[{Clark et~al.(2020)Clark, Luong, Le, and Manning}]{clark2020electra}
Kevin Clark, Minh-Thang Luong, Quoc~V. Le, and Christopher~D. Manning. 2020.
\newblock \href {https://openreview.net/forum?id=r1xMH1BtvB} {Electra:
  Pre-training text encoders as discriminators rather than generators}.
\newblock In \emph{International Conference on Learning Representations
  (ICLR)}.

\bibitem[{Dai and Callan(2019)}]{Dai2019sigir}
Zhuyun Dai and Jamie Callan. 2019.
\newblock \href {https://doi.org/10.1145/3331184.3331303} {Deeper text
  understanding for ir with contextual neural language modeling}.
\newblock \emph{Proceedings of the 42nd International ACM SIGIR Conference on
  Research and Development in Information Retrieval}.

\bibitem[{Devlin et~al.(2019)Devlin, Chang, Lee, and
  Toutanova}]{devlin2019bert}
Jacob Devlin, Ming-Wei Chang, Kenton Lee, and Kristina Toutanova. 2019.
\newblock \href {https://doi.org/10.18653/v1/N19-1423} {{BERT}: Pre-training of
  deep bidirectional transformers for language understanding}.
\newblock In \emph{Proceedings of the 2019 Conference of the North {A}merican
  Chapter of the Association for Computational Linguistics: Human Language
  Technologies, Volume 1 (Long and Short Papers)}, pages 4171--4186,
  Minneapolis, Minnesota. Association for Computational Linguistics.

\bibitem[{Guu et~al.(2020)Guu, Lee, Tung, Pasupat, and Chang}]{guu2020realm}
Kelvin Guu, Kenton Lee, Zora Tung, Panupong Pasupat, and Ming-Wei Chang. 2020.
\newblock Realm: Retrieval-augmented language model pre-training.
\newblock \emph{arXiv preprint arXiv:2002.08909}.

\bibitem[{Hofst{\"a}tter et~al.(2020)Hofst{\"a}tter, Althammer, Schr{\"o}der,
  Sertkan, and Hanbury}]{hofstaetter2020_crossarchitecture_kd}
Sebastian Hofst{\"a}tter, Sophia Althammer, Michael Schr{\"o}der, Mete Sertkan,
  and Allan Hanbury. 2020.
\newblock \href {http://arxiv.org/abs/2010.02666} {Improving efficient neural
  ranking models with cross-architecture knowledge distillation}.

\bibitem[{Humeau et~al.(2020)Humeau, Shuster, Lachaux, and
  Weston}]{Humeau2020Poly-encoders}
Samuel Humeau, Kurt Shuster, Marie-Anne Lachaux, and Jason Weston. 2020.
\newblock \href {https://openreview.net/forum?id=SkxgnnNFvH} {Poly-encoders:
  Architectures and pre-training strategies for fast and accurate
  multi-sentence scoring}.
\newblock In \emph{International Conference on Learning Representations
  (ICLR)}.

\bibitem[{Jiao et~al.(2020)Jiao, Yin, Shang, Jiang, Chen, Li, Wang, and
  Liu}]{tinybert2020jiao}
Xiaoqi Jiao, Yichun Yin, Lifeng Shang, Xin Jiang, Xiao Chen, Linlin Li, Fang
  Wang, and Qun Liu. 2020.
\newblock \href {https://doi.org/10.18653/v1/2020.findings-emnlp.372}
  {{T}iny{BERT}: Distilling {BERT} for natural language understanding}.
\newblock In \emph{Findings of the Association for Computational Linguistics:
  EMNLP 2020}, pages 4163--4174, Online. Association for Computational
  Linguistics.

\bibitem[{Johnson et~al.(2017)Johnson, Douze, and J{\'e}gou}]{2017faiss}
Jeff Johnson, Matthijs Douze, and Herv{\'e} J{\'e}gou. 2017.
\newblock Billion-scale similarity search with gpus.
\newblock \emph{arXiv preprint arXiv:1702.08734}.

\bibitem[{Karpukhin et~al.(2020)Karpukhin, Oguz, Min, Lewis, Wu, Edunov, Chen,
  and Yih}]{karpukhin-etal-2020-dense}
Vladimir Karpukhin, Barlas Oguz, Sewon Min, Patrick Lewis, Ledell Wu, Sergey
  Edunov, Danqi Chen, and Wen-tau Yih. 2020.
\newblock \href {https://doi.org/10.18653/v1/2020.emnlp-main.550} {Dense
  passage retrieval for open-domain question answering}.
\newblock In \emph{Proceedings of the 2020 Conference on Empirical Methods in
  Natural Language Processing (EMNLP)}, pages 6769--6781, Online. Association
  for Computational Linguistics.

\bibitem[{Khattab et~al.(2021)Khattab, Potts, and
  Zaharia}]{khattab2020relevance}
Omar Khattab, Christopher Potts, and Matei Zaharia. 2021.
\newblock \href {https://doi.org/10.1162/tacl_a_00405} {Relevance-guided
  supervision for {O}pen{QA} with {C}ol{BERT}}.
\newblock \emph{Transactions of the Association for Computational Linguistics},
  9:929--944.

\bibitem[{Khattab and Zaharia(2020)}]{khattab2020colbert}
Omar Khattab and Matei Zaharia. 2020.
\newblock Colbert: Efficient and effective passage search via contextualized
  late interaction over bert.
\newblock In \emph{Proceedings of the 43rd International ACM SIGIR Conference
  on Research and Development in Information Retrieval}, pages 39--48.

\bibitem[{Kwiatkowski et~al.(2019)Kwiatkowski, Palomaki, Redfield, Collins,
  Parikh, Alberti, Epstein, Polosukhin, Devlin, Lee, Toutanova, Jones, Kelcey,
  Chang, Dai, Uszkoreit, Le, and Petrov}]{naturalquestions}
Tom Kwiatkowski, Jennimaria Palomaki, Olivia Redfield, Michael Collins, Ankur
  Parikh, Chris Alberti, Danielle Epstein, Illia Polosukhin, Jacob Devlin,
  Kenton Lee, Kristina Toutanova, Llion Jones, Matthew Kelcey, Ming-Wei Chang,
  Andrew~M. Dai, Jakob Uszkoreit, Quoc Le, and Slav Petrov. 2019.
\newblock \href {https://doi.org/10.1162/tacl_a_00276} {Natural questions: A
  benchmark for question answering research}.
\newblock \emph{Transactions of the Association for Computational Linguistics},
  7:452--466.

\bibitem[{Larson et~al.(2019)Larson, Mahendran, Peper, Clarke, Lee, Hill,
  Kummerfeld, Leach, Laurenzano, Tang, and Mars}]{larson-etal-2019-evaluation}
Stefan Larson, Anish Mahendran, Joseph~J. Peper, Christopher Clarke, Andrew
  Lee, Parker Hill, Jonathan~K. Kummerfeld, Kevin Leach, Michael~A. Laurenzano,
  Lingjia Tang, and Jason Mars. 2019.
\newblock \href {https://doi.org/10.18653/v1/D19-1131} {An evaluation dataset
  for intent classification and out-of-scope prediction}.
\newblock In \emph{Proceedings of the 2019 Conference on Empirical Methods in
  Natural Language Processing and the 9th International Joint Conference on
  Natural Language Processing (EMNLP-IJCNLP)}, pages 1311--1316, Hong Kong,
  China. Association for Computational Linguistics.

\bibitem[{Lewis et~al.(2020)Lewis, Perez, Piktus, Petroni, Karpukhin, Goyal,
  K\"{u}ttler, Lewis, Yih, Rockt\"{a}schel, Riedel, and
  Kiela}]{lewis2020retrieval}
Patrick Lewis, Ethan Perez, Aleksandra Piktus, Fabio Petroni, Vladimir
  Karpukhin, Naman Goyal, Heinrich K\"{u}ttler, Mike Lewis, Wen-tau Yih, Tim
  Rockt\"{a}schel, Sebastian Riedel, and Douwe Kiela. 2020.
\newblock \href
  {https://proceedings.neurips.cc/paper/2020/file/6b493230205f780e1bc26945df7481e5-Paper.pdf}
  {Retrieval-augmented generation for knowledge-intensive nlp tasks}.
\newblock In \emph{Advances in Neural Information Processing Systems},
  volume~33, pages 9459--9474. Curran Associates, Inc.

\bibitem[{Lin et~al.(2021)Lin, Yang, and Lin}]{lin-etal-2021-batch}
Sheng-Chieh Lin, Jheng-Hong Yang, and Jimmy Lin. 2021.
\newblock \href {https://doi.org/10.18653/v1/2021.repl4nlp-1.17} {In-batch
  negatives for knowledge distillation with tightly-coupled teachers for dense
  retrieval}.
\newblock In \emph{Proceedings of the 6th Workshop on Representation Learning
  for NLP (RepL4NLP-2021)}, pages 163--173, Online. Association for
  Computational Linguistics.

\bibitem[{Liu et~al.(2019)Liu, Ott, Goyal, Du, Joshi, Chen, Levy, Lewis,
  Zettlemoyer, and Stoyanov}]{roberta2019}
Yinhan Liu, Myle Ott, Naman Goyal, Jingfei Du, Mandar Joshi, Danqi Chen, Omer
  Levy, Mike Lewis, Luke Zettlemoyer, and Veselin Stoyanov. 2019.
\newblock \href {http://arxiv.org/abs/1907.11692} {Roberta: {A} robustly
  optimized {BERT} pretraining approach}.
\newblock \emph{CoRR}, abs/1907.11692.

\bibitem[{Luan et~al.(2021)Luan, Eisenstein, Toutanova, and
  Collins}]{luan-etal-2021-sparse}
Yi~Luan, Jacob Eisenstein, Kristina Toutanova, and Michael Collins. 2021.
\newblock \href {https://doi.org/10.1162/tacl_a_00369} {Sparse, dense, and
  attentional representations for text retrieval}.
\newblock \emph{Transactions of the Association for Computational Linguistics},
  9:329--345.

\bibitem[{MacAvaney et~al.(2020)MacAvaney, Nardini, Perego, Tonellotto,
  Goharian, and Frieder}]{Macavaney2020PreTTR}
Sean MacAvaney, Franco~Maria Nardini, Raffaele Perego, Nicola Tonellotto, Nazli
  Goharian, and Ophir Frieder. 2020.
\newblock \href {https://doi.org/10.1145/3397271.3401093} {Efficient document
  re-ranking for transformers by precomputing term representations}.
\newblock In \emph{Proceedings of the 43rd International ACM SIGIR Conference
  on Research and Development in Information Retrieval}, page 49–58, New
  York, NY, USA. Association for Computing Machinery.

\bibitem[{MacAvaney et~al.(2019)MacAvaney, Yates, Cohan, and
  Goharian}]{MacAvaney2019sigir}
Sean MacAvaney, Andrew Yates, Arman Cohan, and Nazli Goharian. 2019.
\newblock \href {https://doi.org/10.1145/3331184.3331317} {Cedr: Contextualized
  embeddings for document ranking}.
\newblock \emph{Proceedings of the 42nd International ACM SIGIR Conference on
  Research and Development in Information Retrieval}.

\bibitem[{Maillard et~al.(2021)Maillard, Karpukhin, Petroni, Yih, Oguz,
  Stoyanov, and Ghosh}]{Maillard2021multidpr}
Jean Maillard, Vladimir Karpukhin, Fabio Petroni, Wen{-}tau Yih, Barlas Oguz,
  Veselin Stoyanov, and Gargi Ghosh. 2021.
\newblock Multi-task retrieval for knowledge-intensive tasks.
\newblock In \emph{{ACL/IJCNLP} {(1)}}, pages 1098--1111. Association for
  Computational Linguistics.

\bibitem[{Nogueira and Cho(2019)}]{nogueira2019passage}
Rodrigo Nogueira and Kyunghyun Cho. 2019.
\newblock Passage re-ranking with bert.
\newblock \emph{arXiv preprint arXiv:1901.04085}.

\bibitem[{Petroni et~al.(2021)Petroni, Piktus, Fan, Lewis, Yazdani, De~Cao,
  Thorne, Jernite, Karpukhin, Maillard, Plachouras, Rockt{\"a}schel, and
  Riedel}]{petroni2021kilt}
Fabio Petroni, Aleksandra Piktus, Angela Fan, Patrick Lewis, Majid Yazdani,
  Nicola De~Cao, James Thorne, Yacine Jernite, Vladimir Karpukhin, Jean
  Maillard, Vassilis Plachouras, Tim Rockt{\"a}schel, and Sebastian Riedel.
  2021.
\newblock \href {https://doi.org/10.18653/v1/2021.naacl-main.200} {{KILT}: a
  benchmark for knowledge intensive language tasks}.
\newblock In \emph{Proceedings of the 2021 Conference of the North American
  Chapter of the Association for Computational Linguistics: Human Language
  Technologies}, pages 2523--2544, Online. Association for Computational
  Linguistics.

\bibitem[{Qi et~al.(2021)Qi, Pan, Sood, Shah, Kunc, Yu, and
  Potdar}]{qi-etal-2021-benchmarking}
Haode Qi, Lin Pan, Atin Sood, Abhishek Shah, Ladislav Kunc, Mo~Yu, and Saloni
  Potdar. 2021.
\newblock \href {https://doi.org/10.18653/v1/2021.naacl-industry.38}
  {Benchmarking commercial intent detection services with practice-driven
  evaluations}.
\newblock In \emph{Proceedings of the 2021 Conference of the North American
  Chapter of the Association for Computational Linguistics: Human Language
  Technologies: Industry Papers}, pages 304--310, Online. Association for
  Computational Linguistics.

\bibitem[{Qu et~al.(2021)Qu, Ding, Liu, Liu, Ren, Zhao, Dong, Wu, and
  Wang}]{qu-etal-2021-rocketqa}
Yingqi Qu, Yuchen Ding, Jing Liu, Kai Liu, Ruiyang Ren, Wayne~Xin Zhao, Daxiang
  Dong, Hua Wu, and Haifeng Wang. 2021.
\newblock \href {https://doi.org/10.18653/v1/2021.naacl-main.466}
  {{R}ocket{QA}: An optimized training approach to dense passage retrieval for
  open-domain question answering}.
\newblock In \emph{Proceedings of the 2021 Conference of the North American
  Chapter of the Association for Computational Linguistics: Human Language
  Technologies}, pages 5835--5847, Online. Association for Computational
  Linguistics.

\bibitem[{Robertson et~al.(1995)Robertson, Walker, Jones, Hancock-Beaulieu, and
  Gatford}]{bm25}
Stephen~E. Robertson, Steve Walker, Susan Jones, Micheline Hancock-Beaulieu,
  and Mike Gatford. 1995.
\newblock Okapi at {TREC-3}.
\newblock In \emph{Overview of the Third Text REtrieval Conference (TREC-3)},
  pages 109--126. National Institute of Standards and Technology (NIST).

\bibitem[{Sanh et~al.(2019)Sanh, Debut, Chaumond, and
  Wolf}]{sanh2019distilbert}
Victor Sanh, Lysandre Debut, Julien Chaumond, and Thomas Wolf. 2019.
\newblock \href {http://arxiv.org/abs/1910.01108} {{DistilBERT, a distilled
  version of BERT: smaller, faster, cheaper and lighter}}.
\newblock In \emph{5th Workshop on Energy Efficient Machine Learning and
  Cognitive Computing @ NeurIPS 2019}.

\bibitem[{Santhanam et~al.(2021)Santhanam, Khattab, Saad-Falcon, Potts, and
  Zaharia}]{santhanam2021colbertv2}
Keshav Santhanam, Omar Khattab, Jon Saad-Falcon, Christopher Potts, and Matei
  Zaharia. 2021.
\newblock \href {http://arxiv.org/abs/2112.01488} {Colbertv2: Effective and
  efficient retrieval via lightweight late interaction}.

\bibitem[{Simo-Serra et~al.(2015)Simo-Serra, Trulls, Ferraz, Kokkinos, Fua, and
  Moreno-Noguer}]{2015hardnegative}
Edgar Simo-Serra, Eduard Trulls, Luis Ferraz, Iasonas Kokkinos, Pascal Fua, and
  Francesc Moreno-Noguer. 2015.
\newblock Discriminative learning of deep convolutional feature point
  descriptors.
\newblock In \emph{Proceedings of the IEEE International Conference on Computer
  Vision}, pages 118--126.

\bibitem[{Wu et~al.(2017)Wu, Manmatha, Smola, and
  Krähenbühl}]{2017samplingMatters}
Chao-Yuan Wu, R.~Manmatha, Alexander~J. Smola, and Philipp Krähenbühl. 2017.
\newblock \href {https://doi.org/10.1109/ICCV.2017.309} {Sampling matters in
  deep embedding learning}.
\newblock In \emph{2017 IEEE International Conference on Computer Vision
  (ICCV)}, pages 2859--2867.

\bibitem[{Xiong et~al.(2021)Xiong, Xiong, Li, Tang, Liu, Bennett, Ahmed, and
  Overwijk}]{xiong2021ance}
Lee Xiong, Chenyan Xiong, Ye~Li, Kwok-Fung Tang, Jialin Liu, Paul~N. Bennett,
  Junaid Ahmed, and Arnold Overwijk. 2021.
\newblock \href {https://openreview.net/forum?id=zeFrfgyZln} {Approximate
  nearest neighbor negative contrastive learning for dense text retrieval}.
\newblock In \emph{International Conference on Learning Representations}.

\bibitem[{Yu et~al.(2021)Yu, Liu, Xiong, Feng, and Liu}]{yu2021fewshot}
Shi Yu, Zhenghao Liu, Chenyan Xiong, Tao Feng, and Zhiyuan Liu. 2021.
\newblock \href {https://doi.org/10.1145/3404835.3462856} {Few-shot
  conversational dense retrieval}.
\newblock In \emph{Proceedings of the 44th International ACM SIGIR Conference
  on Research and Development in Information Retrieval}, page 829–838, New
  York, NY, USA. Association for Computing Machinery.

\bibitem[{Zhan et~al.(2021)Zhan, Mao, Liu, Guo, Zhang, and
  Ma}]{zhan-2021-hard-neg}
Jingtao Zhan, Jiaxin Mao, Yiqun Liu, Jiafeng Guo, Min Zhang, and Shaoping Ma.
  2021.
\newblock \href {https://doi.org/10.1145/3404835.3462880} {Optimizing dense
  retrieval model training with hard negatives}.
\newblock In \emph{Proceedings of the 44th International ACM SIGIR Conference
  on Research and Development in Information Retrieval}, page 1503–1512, New
  York, NY, USA. Association for Computing Machinery.

\end{thebibliography}
\bibliographystyle{acl_natbib}

\newpage
\appendix
\section{Appendix}
\label{sec:appendix}

\subsection{Hyper-parameters}
Hyper-parameters for ColBERT:
\begin{verbatim}
NQ pre-training batch_size: 192
tuning batch_size: 32
tuning num_epochs: 10
doc_maxlen: 180 
mask-punctuation: true
amp: true
learning_rate: 3e-06
weight_decay: 0.0
adam_eps: 1e-8
similarity: l2
dimension: 128
query_maxlen: 32
doc_maxlen: 128
\end{verbatim}

Hyper-parameters for DPR:
\begin{verbatim}
NQ pre-training batch_size: 144
Full model tuning batch_size: 27
Tiny model tuning batch_size: 80
NQ pre-train warmup_steps: 1237
tuning warmup_steps: 100
NQ pre-train num_train_epochs: 40
tuning num_train_epochs: 100
learning_rate: 2e-5
weight_decay: 0.0
adam_eps: 1e-8
adam_betas: (0.9, 0.999)
max_grad_norm: 2.0
hard_negatives: 1
other_negatives: 0
\end{verbatim}

\subsection{More Results}
Match@3 scores could be found in
Table~\ref{table-results-m3-1} and Table~\ref{table-results-m3-2}.

\begin{table}
	\centering
	\small

	\begin{tabular}{rlccc}
		\hline
		\hline
		 & \textbf{HRFAQ} & \multicolumn{1}{c}{\textbf{0-shot}} & \multicolumn{1}{c}{\textbf{1 ex/doc}} &  \multicolumn{1}{c}{\textbf{3 ex/doc}}   \\ \hline
%		\# & System & M@1    & M@1   & M@1    \\ \hline
		\hline
    1 &BM25   & 41.3  &  - & -   \\
	2 &	NSVM  & -  & 35.2(4.9) & 58.3(3.1)   \\
	3 &	NSVM w/ text  & 18.7 & 42.8(2.5)  & 61.9(1.9)   \\
	4 &	{DPR$_{full}$} & 43.9   & 58.1(2.7)  & 67.2(1.6)   \\
	5 &{ColBERT$_{full}$} & 53.7 & 62.5(1.0) & 67.1(1.8)   \\
	    \hline
	6 &	{DPR$_{tiny}$(S)} & 37.0  & 52.8(1.7) & 61.0(2.3)     \\
	7 &{ColBERT$_{tiny}$} &  45.3  & 56.4(1.8) &  65.2(1.1) \\
	8 & Ensemble(1,7) & \textbf{49.6}  & 59.2(1.5) & 66.8(1.1)  \\
	9 & Ensemble(3,7) & 45.6 & \textbf{60.0(1.8)}  & \textbf{69.1(1.4)}  \\
		\hline
		\hline
	\end{tabular}

	%\vspace{15pt}
\caption{\label{table-results-m3-1}
Match@3 scores on HRFAQ testset.
For $k$ ex/doc experiments: we take 10 random seeds; for each random seed, sample $k$ training
queries per answer text, train a model; finally report avg(std) of the 10 models.
	}
\end{table}

\begin{table}
\centering
\small
\begin{tabular}{rlccc}
	\hline
	\hline
& \textbf{MEDFAQ} & \multicolumn{1}{c}{\textbf{0-shot}} & \multicolumn{1}{c}{\textbf{1 ex/doc}} &  \multicolumn{1}{c}{\textbf{3 ex/doc}}   \\ \hline
%		\# & System & M@1    & M@1   & M@1    \\ \hline
	\hline
    1 &BM25   &  37.2 &  - & -   \\
	2 &	NSVM  & -  & 53.9(7.1) & 68.9(6.1)   \\
	3 &	NSVM w/ text  & 33.5 & 55.5(4.6) & 70.6(6.4)    \\
	4 &	{DPR$_{full}$} & 47.4 & 72.8(2.7)  & 78.7(1.6)   \\
	5 &{ColBERT$_{full}$} & 61.7  & 74.1(2.0)  & 79.8(1.4)  \\
	    \hline
	6 &	{DPR$_{tiny}$(S)} & 35.3 & 56.3(4.7) & 70.7(3.6)    \\
	7 &{ColBERT$_{tiny}$} & 38.5  & 62.7(3.0)  & 73.1(3.0) \\
	8 & Ensemble(1,7) & \textbf{41.8}  & 63.6(3.0)  & 73.8(3.3)   \\
	9 & Ensemble(3,7) & 40.26 & \textbf{63.7(5.8)}  & \textbf{74.4(4.8)}   \\
	\hline
	\hline
\end{tabular}
	%\vspace{15pt}
\caption{\label{table-results-m3-2}
Match@3 scores on MEDFAQ testset.
For $k$ ex/doc experiments: we take 10 random seeds; for each random seed, sample $k$ training
queries per answer text, train a model; finally report avg(std) of the 10 models.
	}
\end{table}

\subsection{Licenses and Potential Risks}

The licenses of ColBERT code and DPR code can be found at
\url{https://github.com/stanford-futuredata/ColBERT/blob/master/LICENSE} and
\url{https://github.com/facebookresearch/DPR/blob/main/LICENSE}, respectively.
The license of ElasticSearch can be found at
\url{https://github.com/elastic/elasticsearch/blob/7.16/licenses/ELASTIC-LICENSE-2.0.txt}.
The neural embedding based SVM classifier is part of commercial products owned by our organization.

We ran the experiments on our own extracted datasets for solely research exploration purpose, and we did not distribute or use the code or data to make any profit. The datasets are small to check / anonymize. We use them solely for benchmarking purpose, and strictly protected access to the datasets to only a couple of co-authors.

Our work is exploring the efficient and effective approaches of text retrieval on answer text corpus curated by chat-bot administrators. The use case is how to present the most matching answer text to users, where the answer text itself is created and closely administered by chat-bot administrators. The scope of this paper does not cover research on how to filter offensive content. On the other hand, our work does not generate any new text, hence does not create risks to users.

\end{document}